\PassOptionsToPackage{prologue,dvipsnames}{xcolor}
\documentclass[a4paper,10pt]{article}
\usepackage{jheppub}

\usepackage{amsmath,amssymb,amsfonts,mathtools}
\usepackage{physics}
\usepackage{bbm}
\usepackage{tensor}
\usepackage{youngtab}
\usepackage{simplewick}
\usepackage{slashed}
\usepackage{cancel}
\usepackage{nicematrix}

\usepackage{graphicx}
\usepackage{caption}
\usepackage{subcaption}
\usepackage{float}
\usepackage{wrapfig}
\usepackage{tikz}
\usetikzlibrary{arrows.meta, positioning, decorations.pathmorphing, shapes}
\usepackage[compat=1.1.0]{tikz-feynman}
\tikzfeynmanset{warn luatex=false}
\usepackage{pgfplots}       
\pgfplotsset{compat=1.18}

\usepackage{booktabs}
\usepackage{multirow}
\usepackage{multicol}
\usepackage{array}
\usepackage{tabularx}

\usepackage[utf8]{inputenc} 
\usepackage[T1]{fontenc}    
\usepackage{microtype}      
\usepackage{xspace}         
\usepackage{textcomp}       
\usepackage{gensymb}        
\usepackage{wasysym}        

\usepackage{setspace}
\usepackage{parskip}
\usepackage{indentfirst}
\usepackage{footmisc}
\usepackage{fancyhdr}
\usepackage{titlesec}
\usepackage{footnote}
\usepackage[colorlinks=true,
            linkcolor=blue,
            citecolor=blue,
            urlcolor=blue]{hyperref}
\usepackage[noabbrev]{cleveref}

\usepackage{xcolor}
\usepackage{orcidlink}
\usepackage{xifthen}        
\usepackage{comment}        
\usepackage{verbatim}       
\usepackage{lipsum}         

\newcommand{\n}{\nonumber}

\renewcommand{\d}{\partial}

\newcommand{\bee}{\begin{eqnarray}}
\newcommand{\eee}{\end{eqnarray}}
\newcommand{\be}{\begin{equation}}
\newcommand{\ee}{\end{equation}}
\newcommand{\benn}{\begin{equation*}}
\newcommand{\eenn}{\end{equation*}}
\newcommand{\ba}{\begin{array}} 
\newcommand{\ea}{\end{array}}
\def\bea{\begin{eqnarray}}
\def\eea{\end{eqnarray}}

\newcommand{\Dfb}{\mathord{\buildrel{\lower3pt\hbox{$\scriptscriptstyle{\leftrightarrow \tiny{ \ \ \ } }$}}\over {D^{\mu}}}}
\newcommand{\Dfbd}{\mathord{\buildrel{\lower3pt\hbox{$\scriptscriptstyle\leftrightarrow$}}\over {D}_{\mu}}}

\renewcommand{\L}{\mathcal{L}}

\begin{document}
\allowdisplaybreaks
\title{
Background Fields Meet the Heat Kernel: Gauge Invariance and RGEs without diagrams}
\abstract{
 We introduce a new method that exploits the combination of the Heat Kernel (HK) and Background Field Method to compute gauge-invariant and gauge parameter-independent quantities such as the effective potential, anomalous dimensions, and renormalization group equations. In contrast to currently employed techniques, these results are obtained {\emph{exclusively}} from the dynamics of the background fields, without relying on supplementary input from, e.g., traditional diagrammatic calculations. This is achieved by a consistent treatment of open and closed derivatives in the HK expansions. In this way, we compute the standard quantities such as $\beta$ functions and their gauge-parameter independence when background fields are on-shell. We demonstrate this formalism for instructive examples such as Scalar QED and Yukawa theory. Full results for the bosonic part of the Standard Model provide further validation of our approach.
}

\author{Debanjan Balui$^{\vartheta}$\orcidlink{0009-0007-0917-2078},}
\author{Joydeep Chakrabortty$^{\vartheta}$\orcidlink{0000-0001-8709-916X},}
\author{Christoph~Englert$^\varsigma$\orcidlink{0000-0003-2201-0667},}
\author{Subhendra Mohanty$^\varphi$\orcidlink{0000-0003-0070-6647},}
\author{Tushar$^{\vartheta}$\orcidlink{0009-0000-4057-7574}}
\affiliation{$^{\vartheta}$Indian Institute of Technology Kanpur, Kalyanpur, Kanpur 208016, Uttar Pradesh, India}
\affiliation{$^{\varsigma}$Department of Physics \& Astronomy, University of Manchester, Manchester M13 9PL, United Kingdom}
\affiliation{$^{\varphi}$IISER Bhopal, Bhopal 462066, Madhya Pradesh, India}
\emailAdd{debanjanb22@iitk.ac.in}
\emailAdd{joydeep@iitk.ac.in}
\emailAdd{subhendram@iiserb.ac.in}
\emailAdd{christoph.englert@manchester.ac.uk}
\emailAdd{tushar25@iitk.ac.in}
\maketitle
\flushbottom
\section{Introduction and Results}
\label{sec:intro}
The Background Field Method (BFM)~\cite{DeWitt:1967ub, Boulware:1980av, Kluberg-Stern:1974nmx, Kluberg-Stern:1975ebk, Hart:1983lbv, Abbott:1981ke, Abbott:1980hw} is a standard tool to compute the one-loop effective potential~\cite{Coleman:1973jx}, but not limited to this application, see e.g.~\cite{Denner:1994nn, Denner:1994xt, Denner:2019vbn} for discussion in the context of radiative corrections in Standard Model (SM) computations. In the BFM, a quantum field is decomposed into background (treated as a classical source) and its quantum fluctuations (the relevant integration parameter of the generating functional), and the effective potential is computed after integrating out those fluctuations. In the case of a gauge theory, along with the Background Gauge Fixing (BGF) term, this method ensures that all the radiatively generated operators are invariant under background gauge transformation. This leads to many theoretically desirable features.

In parallel, the gauge-dependence of radiatively-corrected quantities is a relevant issue. Whilst minima of the effective potential are gauge-independent via the Nielsen identities~\cite{Nielsen:1975fs}, excursions away from the minima can be gauge-dependent. This requires careful consideration of the question at hand, see, e.g.,~\cite{Andreassen:2014gha,  DiLuzio:2014bua, Espinosa:2015qea}. In previous works \cite{Balui:2025kat, Balui:2025yvd}, it has been noted that either utilizing the so-called multiplicative anomaly~\cite{Elizalde:1997nd, Elizalde:1998xq, Bytsenko:2003tu} or the elliptic matrix operator in the Heat Kernel (HK) method, one can compute manifestly gauge parameter-independent effective potentials.  The HK method has been extensively used for background gauge theories in \cite{Jack:1982hf,Jack:1982sr,Jack:1984vj,tHooft:1973bhk,Ichinose:1981uw}.

The naive implementation of the BFM fails to produce the anomalous dimensions (and therefore renormalization group equations, RGEs) of the parameters of the theory. This hampers the self-consistent RGE-improvement of effective potentials~\cite{Coleman:1973jx}, which becomes relevant for cosmological phase transitions~\cite{Quiros:1994dr, Quiros:1999jp, Masina:2025pnp, Chakrabortty:2024wto}. The general practice here is to compute the effective potential using the BFM, but borrow the anomalous dimensions of the fields from a Feynman diagram-based approach. In this work, we introduce a self-consistent approach, using the BFM, that enables the computation of gauge-independent effective potential {\emph{as well as}} RGEs. The procedure outlined in this work gives rise to the anomalous dimensions of fields (and related quantities) as gauge parameter-independent quantities when we invoke the on-shell criteria of the background fields. 
\vspace{0.275cm}

\begin{figure}[hbt!]
    \centering
    \includegraphics[width=\linewidth]{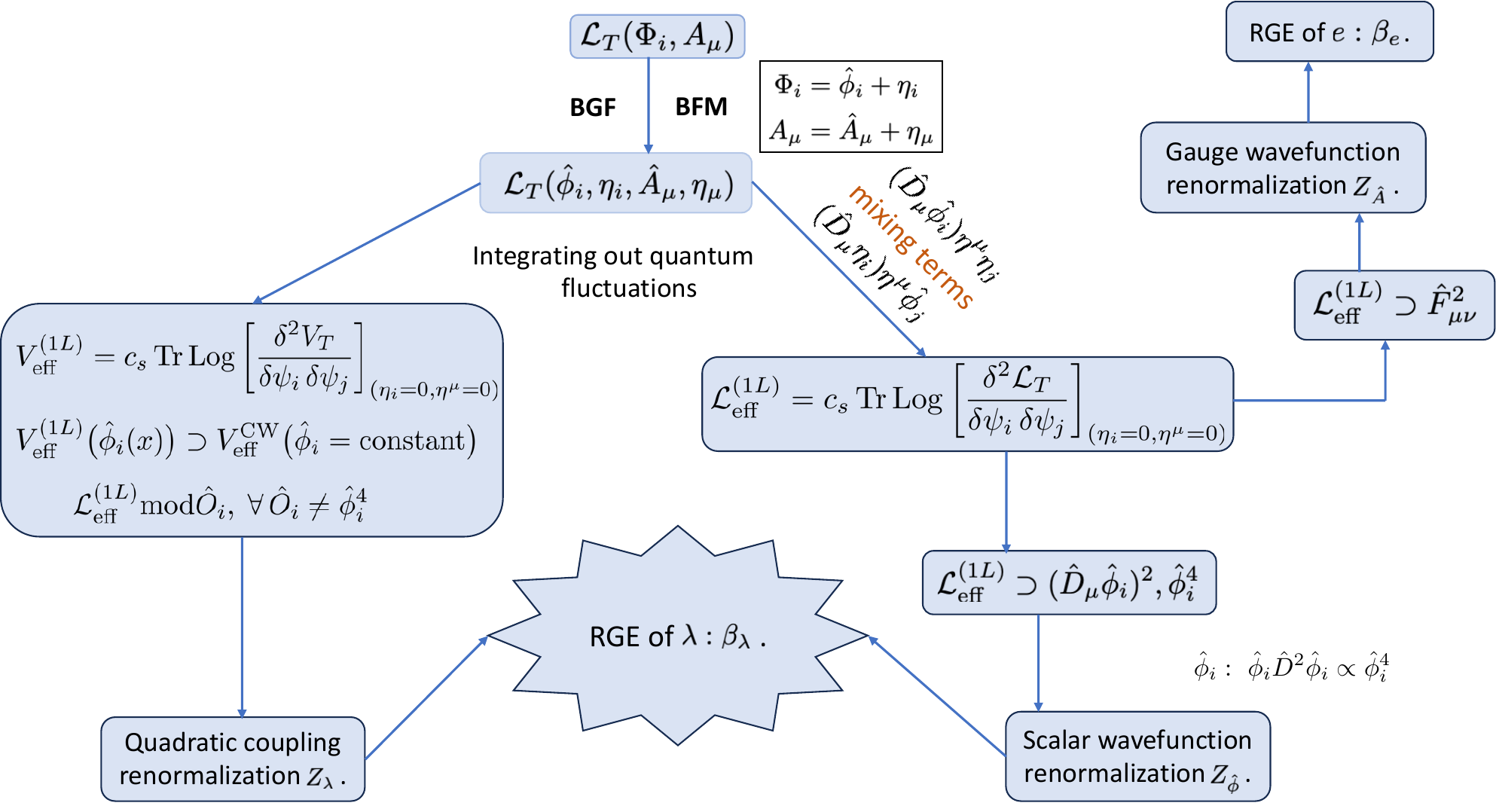}
    \caption{Outline of the strategy to obtain anomalous dimensions in the BFM, starting with a tree-level Lagrangian ${\cal{L}}_T$ to obtain the one-loop effective Lagrangian ${\cal{L}}_\text{eff}^{(1L)}$. Integration-by-parts, renormalization, and the application of background-field equations of motion reproduce the $\beta$ functions in the diagrammatic approach, specifically in the less trivial case of a non-gauge coupling $\lambda$ (we detail the examples of MSQED and Yukawa theory in Sec.~\ref{sec:ex}).}
    \label{fig:SQED-flowchart}
\end{figure}

Our methodology can be summarized as follows (see also Fig.~\ref{fig:SQED-flowchart}):
\begin{itemize}
    \item We start with a tree-level Lagrangian ${\cal{L}}_T$ for given matter fields within a gauge theory.
    \item We expand all the quantum fields into background and quantum fluctuations; BGF ensures the gauge-invariance at every step of the computation and also prohibits the mixing of gauge-invariant operators with non-invariant ones. 
    \item We integrate out the quantum fluctuations and compute the one-loop gauge-invariant effective Lagrangian employing the HK method. Gauge-invariance, understood as invariance of the classical Lagrangian under gauge transformations leading to Ward identities, is obtained through the application of the BFM. Note that gauge parameter-independence of the QFT's building blocks is not directly guaranteed.\footnote{This is typically not considered as alarming as, e.g., Lagrangian parameters do not directly correspond to physical observations such as scattering probabilities; although there can be bookkeeping issues when mixing different renormalization schemes~\cite{Denner:2019vbn}. The gauge parameter-dependence becomes problematic when phenomenological consequences are drawn from interpolating quantities such as the effective potential, cf. the discussion of~\cite{DiLuzio:2014bua}.} 
    \item Focusing on the interaction term (both finite and divergent) $\propto \hat\phi^4$, we identify the counter term $Z_\lambda$ from the divergent piece of the effective potential. We note that, assuming the background fields to be constant, we recover the gauge-invariant Coleman-Weinberg effective potential. 
    \item We consider the background fields to be on-shell; they are not constants. Their on-shell equation of motion schematically reads $\hat\phi \Box \hat\phi \propto \hat\phi^4 $. The mixing of background and quantum fluctuation occurs at the tree-level Lagrangian, and they lead to additional counterterms proportional to $|D_\mu \hat\phi|^2$ and $\hat\phi^4$. After employing equations of motion (EOMs) for the background fields, we find the effective counterterms associated with $|D_\mu \hat\phi|^2$. This leads to the anomalous dimensions $Z_{\hat\phi}$ of the background fields. 
    \item The $Z_{\hat\phi}$, is gauge-invariant and so the anomalous dimension $\gamma_{\hat \phi}$. Here, the effective potential and the $Z_\lambda$ are both gauge-invariant. These together, by construction, lead to the gauge-invariant $\beta$ functions of the theory that are in agreement with a more traditional Feynman diagram-based approach. 
\end{itemize}

Throughout this work, we rely on the HK method (in the BFM with BGF) to compute effective potential, anomalous dimensions, and RGEs. We detail the cancellation of gauge parameters from these quantities when the background fields are on-shell. We also discuss the off-shell regime and the implications of the RGEs for it; an extension of the formalism to Yukawa interactions and to the Standard Model is provided as further examples.

Our work is organized as follows: Section~\ref{sec:ex} is devoted to instructive examples. In Sec.~\ref{sec:MSQEDADBF}, we illustrate the effective Lagrangian computation using the Heat Kernel method relying on BFM along with BGF. We choose Massive Scalar QED as a representative example. We note the shortcomings of this method in the computation of the anomalous dimension of the scalar field that leads to the wrong RGEs of the theory.  In Sec.~\ref{subsec:Ano_Dim_Open_Closed_Der}, we shed light on the missing and previously ignored part of the background-expanded Lagrangian. In the follow-up sections, we elaborate on the notion of closed and open derivatives to capture the {\it missing} contribution due to mixing between background and quantum fluctuations of the scalar fields. We show explicitly that our modified proposal enables the computation of the effective potential, anomalous dimensions, and the RGEs that are gauge-invariant and gauge parameter-independent, too. In Sec.~\ref{sec: Yukawa}, we validate our method for the massive Yukawa theory, where non-trivial anomalous dimensions emerge due to the interaction between scalar and fermion fields. In the appendix, we summarize the results for the bosonic part of the SM and further establish the robustness of our proposal.

\section{Example Case Studies: Massive Scalar QED and Yukawa Theory}
\label{sec:ex}
\subsection{Massive Scalar QED}
\label{sec:MSQEDADBF}
To illustrate and develop the method outlined in the introduction and Fig.~\ref{fig:SQED-flowchart}, it is convenient to consider a simplified toy setting first, before turning to more realistic field theories such as the SM (results for the latter are relegated to an appendix). To this end, we consider Massive Scalar Quantum Electrodynamics (MSQED), which provides a suitable minimal framework for our approach. The MSQED Lagrangian is given by \cite{Dolan:1974gu, Kang:1974yj, Andreassen:2014eha}
\bea \label{eq:MSQEDL}
    \L = \underbrace{-\frac{1}{4} F_{\mu \nu} F^{\mu \nu} + \frac{1}{2} M^2 A^2 - \frac{1}{2\xi} (\partial_\mu A^\mu)^2}_{=\L_\text{gauge}} \n    \\& \hspace{-2cm}
    + \underbrace{|D_\mu \Phi|^2 - M^2 (\Phi^{\dagger} \Phi) - \frac{\lambda}{6} (\Phi^{\dagger} \Phi)^2}_{=\L_\text{scalar}} \;,
\eea
where $\Phi =  (\phi_1 + i \phi_2)/\sqrt{2}$ is the complex scalar field. The covariant derivative is defined as $D_\mu \equiv \partial_\mu + i e A_\mu$, with $A_\mu$ being the $U(1)$ gauge field. Here, the mass terms $\sim M^2$ for scalar and gauge fields are assumed to employ the HK method consistently; they act as infrared regulators. Therefore, these parameters do not play a role in the $\beta$ function calculation of marginal operators that govern the UV dynamics.

\subsubsection{Background Field Expansion and Effective Action} 
\label{subsec:MSQEDQL}
The first step, then, is to apply the BFM. We start with the scalar sector and split the scalar field into classical background and quantum contributions $\Phi=\hat \phi + \eta$. Throughout this work, hatted quantities such as $\hat\phi$ refer to contributions only from the background. This scalar background satisfies the classical EOM. In the same spirit, the gauge field $A_\mu$ is divided into $\hat{A}_\mu$ and the quantum field $\eta_\mu$. Our results depend on the parts that are quadratic in the quantum fields, which are readily obtained through
\bea \label{eq:MSQEDSL}
    \L_\text{scalar}
    &=& 
        (D_\mu \Phi)^{\dagger} (D^\mu \Phi) - M^2 (\Phi^{\dagger} \Phi) - \frac{\lambda}{6} (\Phi^{\dagger} \Phi)^2
        \n \\  
    &\hspace{-1.8 cm} \supset& \hspace{-1 cm}
        \frac{1}{2} \Big\{(\hat{D}_\mu \eta_1) (\hat{D}^\mu \eta_1) 
        - (M^2 + \frac{\lambda}{2} \hat\phi_1^2 + \frac{\lambda}{6} \hat\phi_2^2) \eta_1^2\Big\} 
        + \frac{1}{2} \Big\{(\hat{D}_\mu \eta_2) (\hat{D}^\mu \eta_2) - (M^2 + \frac{\lambda}{2} \hat\phi_2^2 + \frac{\lambda}{6} \hat\phi_1^2) \eta_2^2\Big\} \n \\
    && \hspace{-1 cm}
        + \frac{1}{2} \Big\{2e (\hat{D}_\mu \hat\phi_2) \eta^\mu \eta_1 - 2e (\hat D_\mu \eta_1) \eta^\mu \hat \phi_2 \Big \} - \frac{1}{2} \Big \{2e (\hat D_\mu \hat \phi_1) \eta^\mu \eta_2 - 2e (\hat D_\mu \eta_2) \eta^\mu \hat \phi_1 \Big \} \n \\
    && \hspace{-1 cm}    
        - \frac{\lambda}{3} \hat\phi_1 \hat\phi_2 \eta_1 \eta_2 + \frac{e^2}{2} \hat\phi^2 \eta_\mu \eta^\mu \n \\ 
    &\hspace{-1.8 cm} =& \hspace{-1 cm}
        \textcolor{red}{
        -\frac{1}{2} \Big\{\eta_1 \hat{D}^2 \eta_1 + (M^2 + \frac{\lambda}{2} \hat\phi_1^2 + \frac{\lambda}{6} \hat\phi_2^2) \eta_1^2\Big\} 
        - \frac{1}{2} \Big\{\eta_2 \hat D^2 \eta_2 + (M^2 + \frac{\lambda}{2} \hat\phi_2^2 + \frac{\lambda}{6} \hat\phi_1^2) \eta_2^2\Big\} 
        - \frac{\lambda}{3} \hat\phi_1 \hat\phi_2 \eta_1 \eta_2
        } \n \\ 
    && \hspace{-1 cm}
        \textcolor{blue}{
        +\frac{1}{2} \Big\{2e (\hat D_\mu \hat\phi_2) \eta^\mu 
        \eta_1 - 2e (\hat{D}_\mu \eta_1) \eta^\mu \hat\phi_2\Big\} 
        -\frac{1}{2} \Big\{2e (\hat{D}_\mu \hat\phi_1) \eta^\mu \eta_2 - 2e (\hat D_\mu \eta_2) \eta^\mu \hat \phi_1\Big\}
        } \n \\
    && \hspace{-1 cm}
        + \textcolor{magenta}{\frac{e^2}{2} \hat\phi^2 \eta_\mu \eta^\mu} \;.
\eea
The part highlighted in red (the third line from the bottom) gives the pure scalar part. The blue section (second line from the bottom) corresponds to the scalar-gauge interaction contribution. It is practical to treat the last term as part of $\L_\text{gauge}$, and we define, for convenience,
\bea \label{eq:MSQEDPSL}
    \L_{\substack{\text{scalar} \\ \text{kinetic}}} 
    &=&
        \textcolor{red}{
        -\frac{1}{2} \Big\{\eta_1 \hat{D}^2 \eta_1 + (M^2 + \frac{\lambda}{2} \hat\phi_1^2 + \frac{\lambda}{6} \hat\phi_2^2) \eta_1^2\Big\} 
        - \frac{1}{2} \Big\{\eta_2 \hat{D}^2 \eta_2 + (M^2 + \frac{\lambda}{2} \hat\phi_2^2 + \frac{\lambda}{6} \hat\phi_1^2) \eta_2^2\Big\}} \n \\
    &&
        \textcolor{red}{
        -\frac{\lambda}{3} \hat\phi_1 \hat\phi_2 \eta_1 \eta_2} \;, \\
    \L_{\substack{\text{scalar gauge} \\ \text{interaction}}} 
    &=&
        \textcolor{blue}{
        \frac{1}{2} \Big \{2e (\hat{D}_\mu \hat\phi_2) \eta^\mu \eta_1 - 2e (\hat{D}_\mu \eta_1) \eta^\mu \hat\phi_2\Big\}
        -\frac{1}{2} \Big\{2e (\hat{D}_\mu \hat\phi_1) \eta^\mu \eta_2 - 2e (\hat{D}_\mu \eta_2) \eta^\mu \hat\phi_1\Big\}
        } \n \label{eq:MSQEDLSGIL} \;, \\
    \L_{\substack{\text{gauge} \\ \text{kinetic}}} 
    &=&  
        -\frac{1}{4} F_{\mu \nu} F^{\mu \nu} + \frac{1}{2} M^2 A^2 - \frac{1}{2\xi} (\hat{D}^G_\mu \eta^\mu)^2 + \color{magenta}{\frac{e^2}{2} \hat\phi^2 \eta_\mu \eta^\mu} \n \\
    &=&
        \textcolor{magenta}{
        \frac{1}{2} \eta^\mu \Big\{(\hat{D}_G^2 + M^2 + e^2 \hat\phi^2) g_{\mu \nu} - (1 - \frac{1}{\xi}) \hat{D}^G_\mu \hat{D}^G_\nu\Big\} \eta^\nu} + \dots \label{eq:MSQEDGL} \;.
\eea
For the MSQED, we have $\hat{D}_\mu \equiv \partial_\mu + ie\hat{A}_\mu$, and $\hat{D}^G_\mu \equiv  \d_\mu$. Here, we define the BGF term as $\frac{1}{2\xi} (\hat{D}^G_\mu \eta^\mu)^2$.

A range of techniques is detailed in the literature to obtain the effective action for a given QFT~\cite{Coleman:1973jx, Dolan:1974gu, Vassilevich:2003xt}. Here, we employ the HK method~\cite{  Seeley:1967ea, DeWitt:1964mxt,Belkov:1995gjw, Kirsten:2001wz, Avramidi:2001ns,Vassilevich:2003xt, avramidi2015heat, Avramidi:1990je, Avramidi:1994fx, Banerjee:2023iiv, Banerjee:2023xak, Chakrabortty:2023yke, Banerjee:2024rbc}. This method is rooted in the observation that, when we wish to integrate out heavy states $\Psi$, the two-point contributions take on the form of elliptic operators 
\be \label{eq:ell}
    \Delta_\mathcal{U} \equiv \frac{\delta^2 \L}{\delta \Psi^\dagger \delta \Psi} = \mathcal{\hat{D}}^2 + \mathcal{M}^2 + \mathcal{U} \;,
\ee

with $\mathcal{M}$ denoting the mass of the fields that appear in the loops and $\mathcal{U}$ capturing interactions among them. The effective action at one-loop is determined by~\cite{Vassilevich:2003xt,Banerjee:2023xak}
\vspace{0.05cm}
\be
    \Tr \log \Delta_\mathcal{U} = -\int_0^\infty {\text{d} t\over t} \Tr \exp \{-t \Delta_\mathcal{U}\} + \dots \;,
\ee
where irrelevant constant terms are neglected. All crucial information, therefore, rests in matrix elements
\vspace{0.05cm}
\be \label{eq:delta}
    \left \langle y | \log \Delta_\mathcal{U} | x \right \rangle = 
    -\int_0^\infty {\text{d} t\over t} \, 
    \left \langle y | \exp \{-t \Delta_\mathcal{U}\} | x \right \rangle \equiv -\int_0^\infty {\text{d} t\over t} \, K(t, x, y, \Delta_\mathcal{U}) \;.
\ee
The HK $K(t, x, y, \Delta_\mathcal{U})$ defined above is a solution of the heat equation~\cite{Vassilevich:2003xt,Banerjee:2023xak}
\vspace{0.05cm}
\be \label{eq:HeatEq}
    (\d_t + \Delta_\mathcal{U}) K(t, x, y, \Delta_\mathcal{U}) = 0 \;,
\ee
and it directly obeys the initial condition
\vspace{0.05cm}
\be \label{eq:HeatBC}
    K(0, x, y, \Delta_\mathcal{U}) = \left \langle y | x \right \rangle = \delta(x - y) \;.
\ee
In the free case $\Delta_0 = \d_\mu \d_\mu + M^2$, the solution of Eq.~\eqref{eq:HeatEq} in $d$ dimensions is given by~\cite{Banerjee:2023iiv} 
\vspace{0.05cm}
\be \label{eq:FreeHK}
    K(t, x, y,\Delta_0) = \frac{1}{(4\pi t)^{d/2}}\, \exp \! \Big[\frac{z^2}{4t} - t M^2\Big] \;,
\ee
where $z_\mu=(x - y)_\mu$ (we use an all negative metric convention). The coincidence limit $z \to 0$ required for the computation of the effective action has an asymptotic expansion in $t$ in terms of the so-called Seeley-DeWitt coefficients $b_n$ in the general case $\mathcal{U} \neq 0$
\vspace{0.05cm}
\be
    K(t, x, x, \Delta_\mathcal{U}) = 
    \frac{e^{-t M^2}}{(4\pi t)^{d/2}} \sum_{n=0}^\infty
    b_n(x) \, \frac{(-t)^n}{n!} \;.
\ee  
in $d$ dimensions for a degenerate spectrum. The integration in $t$ (see Eq.~\eqref{eq:delta}) requires a careful regularization~\cite{Balui:2025kat} and leads to an expansion in powers of $M^{-2}$~\cite{Banerjee:2023iiv,Banerjee:2023xak}. This is the desired systematic EFT expansion in the background fields as local, higher-dimensional operators. This structure is made transparent in the free case of Eq.~\eqref{eq:FreeHK}: relevant `time' scales are $t\ll M^{-2}$, where contributions arise dominantly at short distances $z^2\ll t$.

The HK approach can readily be extended to the BFM, as the elliptic operator can be constructed after integrating out the quantum fluctuations. Returning to our MSQED example, we define a multiplet of quantum fluctuations $\Psi= (\eta_1, \eta_2, \eta_\mu)^T$, and identify the elliptic operator as 
\be \label{eq:MSQEDEO}
    \Delta \equiv \\
    \begin{pNiceMatrix}
        \begin{aligned}[c]
            & 
            \hspace{0.5 cm} \hat{D}^2 + M^2 \\
            & 
            + \dfrac{\lambda}{2} \hat\phi_1^2 + \dfrac{\lambda}{6} \hat\phi_2^2 
        \end{aligned}
        \hspace{0.25 cm}
        &
        {\dfrac{\lambda}{3} \hat\phi_1 \hat\phi_2}
        &
        - e (\hat{D}_\nu \hat\phi_2)
        \\[0.6 cm] 
        \hdottedline
        \\[-0.25 cm]
        \dfrac{\lambda}{3} \hat\phi_1 \hat\phi_2
        &
        \hspace{0.25 cm}
        \begin{aligned}[c]
            & 
            \hspace{0.5 cm} \hat{D}^2 + M^2 \\
            & 
            + \dfrac{\lambda}{2} \hat\phi_2^2 + \dfrac{\lambda}{6} \hat\phi_1^2 
        \end{aligned}
        \hspace{0.25 cm}
        &
        e (\hat{D}_\nu \hat\phi_1)
        \\[0.6 cm]
        \hdottedline
        \\[-0.25 cm]
        \begin{aligned}[c]
            - e (\hat{D}_\mu \hat\phi_2)
        \end{aligned}
        &
        \begin{aligned}[c]
            e (\hat{D}_\mu \hat\phi_1)
        \end{aligned}
        &
        \begin{aligned}[c]
            & 
            - (\hat{D}_G^2 + M^2 + e^2 \hat\phi^2) g_{\mu \nu} \\
            & 
            \quad + \left(1 - \xi^{-1}\right) \hat{D}^G_\mu \hat{D}^G_\nu
        \end{aligned}
        \CodeAfter
        \tikz \draw[dashed] (1-|2) -- (6-|2);
        \tikz \draw[dashed] (1-|3) -- (6-|3);
    \end{pNiceMatrix} \;,
\ee
where we ignore the terms containing derivatives of the fluctuations. The canonical form of the elliptic operator 
\be
    \Delta_{i j} \equiv \frac{\delta^2 \L}{\delta \Psi_i \delta \Psi_j} = \mathcal{\hat{D}}^2 + \mathcal{M}^2 + \mathcal{U} \;,
\ee
 is obtained with 
\bea
    &\mathcal{\hat{D}}^2& \equiv 
    \begin{pmatrix}
        \hat{D}^2 \hspace{0.5 cm}
        & 
        \quad 0 \hspace{0.5 cm}
        & 
        \quad 0 
        \\[0.2 cm]
        0 \hspace{0.5 cm}
        & 
        \quad \hat{D}^2 \hspace{0.5 cm}
        & 
        \quad 0
        \\[0.2 cm]
        0 \hspace{0.5 cm}
        &
        \quad 0 \hspace{0.5 cm}
        &
        \quad - \hat{D}_G^2 \, g_{\mu \nu}
    \end{pmatrix} \;,
    \quad
    \mathcal{M}^2 \equiv 
    \begin{pmatrix}
        M^2 \hspace{0.5 cm}
        &
        \quad 0 \hspace{0.5 cm}
        &
        \quad 0
        \\[0.2 cm]
        0 \hspace{0.5 cm}
        &
        \quad M^2 \hspace{0.5 cm}
        &
        \quad 0
        \\[0.2 cm]
        0 \hspace{0.5 cm}
        & 
        \quad 0 \hspace{0.5 cm}
        &
        \quad - M^2\, g_{\mu \nu}
    \end{pmatrix} \;, \n
    \\[0.5 cm]
    &\mathcal{U}& \equiv
    \begin{pNiceMatrix}
        \hspace{0.5 cm} \dfrac{\lambda}{2} \hat\phi_1^2 + \dfrac{\lambda}{6} \hat\phi_2^2 \hspace{0.5 cm}
        &
        \dfrac{\lambda}{3} \hat\phi_1 \hat\phi_2
        &
        - e (\hat{D}_\nu \hat \phi_2)
        \\[0.25cm]
        \hdottedline
        \\[-0.25cm]
        \dfrac{\lambda}{3} \hat\phi_2 \hat\phi_1
        &
        \hspace{0.5 cm} \dfrac{\lambda}{2} \hat\phi_2^2 + \dfrac{\lambda}{6} \hat\phi_1^2 \hspace{0.5 cm}
        &
        e (\hat{D}_\nu \hat\phi_1)
        \\[0.25 cm]
        \hdottedline
        \\[-0.25 cm]
        \begin{aligned}[c]
            - e (\hat{D}_\mu \hat\phi_2)
        \end{aligned}
        &
        \begin{aligned}[c]
            e (\hat{D}_\mu \hat\phi_1)
        \end{aligned}
        &
        \begin{aligned}[c]
            &
            \hspace{0.75 cm} - e^2 \hat\phi^2 g_{\mu \nu} \\
            &
            + \left(1 - \xi^{-1}\right) \hat{D}^G_\mu \hat{D}^G_\nu
        \end{aligned}
        \CodeAfter
        \tikz \draw[dashed] (1-|2) -- (6-|2);
        \tikz \draw[dashed] (1-|3) -- (6-|3);
    \end{pNiceMatrix} \;.
\eea

With the elliptic operator in place, we can now proceed to determine the effective action as outlined above. Ref.~\cite{Balui:2025kat} demonstrated that the effective potential and the other relevant terms of the one-loop effective Lagrangian for the Abelian Higgs model are gauge-invariant and independent of the gauge parameter in this approach. As derived in Ref.~\cite{Balui:2025kat}, we find the one-loop finite as well as divergent contribution as (in dimensional regularization $d=4-2\epsilon$ which implies $e\to e\mu^\epsilon$ and $\lambda \to \lambda\mu^{2\epsilon}$)
\bea \label{eq:EffLb2divMSQEDHK}
  \L_{\text{eff}}^{\text{fin}} + \L_{\text{eff}}^{\text{div}} 
  &=& 
    -\frac{1}{64\pi^2} \left[\frac{5\lambda^2 {\hat\phi}^4}{18} + 3e^4 \hat\phi^4 + 2e^2 (\hat{D}_\mu \hat\phi_a) (\hat{D}_\mu \hat\phi_a) + \frac{e^2}{3} (\hat{F}_{\mu \nu})^2\right] \log \left(\dfrac{M^2}{\mu^2}\right)
    \n \\
  && 
    +\frac{1}{64\pi^2} \frac{1}{\epsilon} \left[\frac{5\lambda^2 \hat\phi^4}{18} + 3e^4 \hat\phi^4 + 2e^2 (\hat{D}_\mu \hat\phi_a) (\hat{D}_\mu \hat\phi_a) 
    + \frac{e^2}{3} (\hat{F}_{\mu \nu})^2\right] \;,
\eea
where, $\hat{F}_{\mu \nu} = [\hat{D}_\mu, \hat{D}_\nu]$, and $|\hat \phi| = (\hat\phi_1^2+\hat\phi_2^2)^{1/2}$.

We renormalize the bare effective Lagrangian in the usual way. The wavefunction renormalization for the scalar field and the quartic coupling are 
\be \label{eq:ren_coef}
    \hat\phi_a^B = Z_{\hat\phi}^{1/2} \hat\phi_a \;, 
    \quad \hat{A}_\mu^B = Z_{\hat{A}}^{1/2} \hat{A}_\mu \;, 
    \quad \lambda^B = \frac{Z_\lambda}{Z_{\hat\phi}^2} \lambda \;, \quad e^B = \frac{Z_e}{Z_{\hat\phi} Z_{\hat{A}}^{1/2}} e \;,
\ee
with
\be \label{eq:ren_coef_div}
    Z_{\hat\phi} = 1 + \frac{1}{\epsilon} \delta Z_{\hat\phi} \;, 
    \quad Z_{\hat{A}} = 1 + \frac{1}{\epsilon} \delta Z_{\hat{A}} \;,
    \quad Z_\lambda = 1 + \frac{1}{\epsilon} \delta Z_\lambda \;, \quad Z_e = 1 + \frac{1}{\epsilon} \delta Z_e \;
\ee
at one loop.\footnote{Note that for this choice of renormalization constants, $Z_e$ refers to the \emph{operator renormalization} and not to the renormalization of the coupling constant directly. Indeed, the latter is straightforwardly recovered from the Ward identity of Eq.~\eqref{eq:coefs}, $Z_{\hat \phi}=Z_{e}$, so that $e_B=Z_{\hat{A}}^{-1/2}e$ as expected.} The UV-divergent part of Eq.~\eqref{eq:EffLb2divMSQEDHK} at one-loop order are
\be \label{eq:coefs}
    \delta Z_{\hat\phi} = \delta Z_e = \frac{1}{16\pi^2} e^2 \;, \quad \delta Z_\lambda = \frac{1}{16\pi^2} 
    \left[\frac{5}{3} \lambda + 18 \frac{e^4}{\lambda}\right] \;,
    \quad \delta Z_{\hat A} = -\frac{1}{16\pi^2} \frac{e^2}{3} \;.
\ee
These results of the previous section enable us to compute the $\beta$ functions of this theory.

\subsubsection{Running couplings}
%
We will now derive the RGE evolution of the parameters of the theory of Eq.~\eqref{eq:MSQEDL} at one-loop order. The regularization parameter in Eq.~\eqref{eq:ren_coef} $\epsilon$ can be understood as the generalized $\overline{\text{MS}}$ factor

\be
    \epsilon^{-1} \to \left(4\pi \mu^2 \over \mu_R^2\right)^{{2 - d/2}} {2 \, \Gamma(3 - d/2) \over 4 - d} \;,
\ee
which introduces the renormalization scale $\mu_R$ that effectively replaces the `t Hooft mass $\mu\to \mu_R$ via the counterterms. Calculations of RGEs can therefore be abridged by investigating their behavior under changes of $\mu$ directly. The $\beta$ function for an arbitrary renormalized coupling $C$ is defined as (see \cite{Bohm:2001yx} for an excellent pedagogical introduction)
\be
    \beta_C = \lim_{\epsilon \to 0} \mu \frac{\d C}{\d \mu} \;.
\ee
In the following, we will treat the limit $d \to 4$ implicitly. Suppose the relation between the bare and renormalized quantity is 
\be
    C^B = \Pi_{i = 1}^N Z_{\psi^i}^{n^i} \, C \, (1 + \dfrac{1}{\epsilon} \delta Z_C) \, \mu^{\Omega_C \epsilon} \;,
\ee
where 
\be
    Z_{\psi^i}^{n^i} = 1 + \frac{1}{\epsilon} n^i \delta Z_{\psi^i} \;.
\ee
The introduction of $\mu^\Omega_C$ is necessary in the case of dimensional regularization to make the renormalized quantity $C$ dimensionless.
\vspace{-0.05cm}

Given the above assumptions, the $\beta$ function takes the form
\bea \label{eq:beta_form}
    \beta_C = \!\!\!
    &&
        C \left[\Omega_C C \dfrac{\d  \delta Z_C}{\d C}  + \sum_j \Omega_{E^j} E^j \dfrac{\d \delta Z_C}{\d \, E^j}\right] 
        \n \\
    && \hspace{2 cm}
        + C \sum_{i = 1}^N n^i \left[\Omega_C \dfrac{\d  \delta Z_{\psi^i}}{\d C} C + \sum_j \Omega_{E^j} E^j\dfrac{\d \delta Z_{\psi^i}}{\d {E^j}}\right] \;,
\eea
where $E$ is a set of other parameters such as the couplings $\{e, y, \dots\}$.

\subsubsection{RGEs from naive effective action calculations}
Using the above formula, we get $\beta_\lambda$ as
\bea
    \beta_\lambda
    &=&
        (-2) \left\{2 \frac{\d\delta Z_{\hat\phi}}{\d\lambda} \, \lambda + \frac{\d\delta Z_{\hat\phi}}{\d e} 
        \, e\right\} \lambda + \lambda \left\{2 \frac{\d\delta Z_\lambda}{\d\lambda} \, \lambda + e \frac{\d\delta Z_\lambda}{\d e}
        \right\}
        \n \\ \label{eq:naive}
    &=&
        \frac{1}{16\pi^2} \left[{-4\lambda e^2} + \frac{10}{3} \lambda^2 + 36e^4\right] \;.
\eea
Similarly, we find the running of $e$ that reads
\benn
    \beta_e = e \left[\frac{\d\delta Z_e}{\d e} e\right] + \left(-\frac{1}{2}\right) \left[\frac{\d\delta Z_{\hat{A}}}{\d e} e\right] e + (-1) \left[\frac{\d \delta Z_{\hat\phi}}{\d e} e\right] e \;,
\eenn
where the Ward identity determines 
\be
    \beta_e = \frac{1}{16\pi^2} \frac{e^3}{3} \;.
\ee
Indeed, this is the correct $\beta$ function for the renormalization group evolution of a coupling constant in the presence of a single, charged scalar particle~\cite{Kang:1974yj,Andreassen:2014eha}. This is a standard result of the BFM as it implements gauge-invariance of the background fields, implying the non-renormalization of $e\hat{A}$, which has been employed to compute the two-loop $\beta$ function~\cite{Abbott:1980hw,Abbott:1983zw} purely from two-point functions in stark contrast to conventional approaches.

\subsubsection{Comparison with the diagrammatic approach} \label{subsec:Diag_Ano_Dim} 
\begin{figure}[hbt!]
    \centering
    \includegraphics[width=0.9\textwidth]{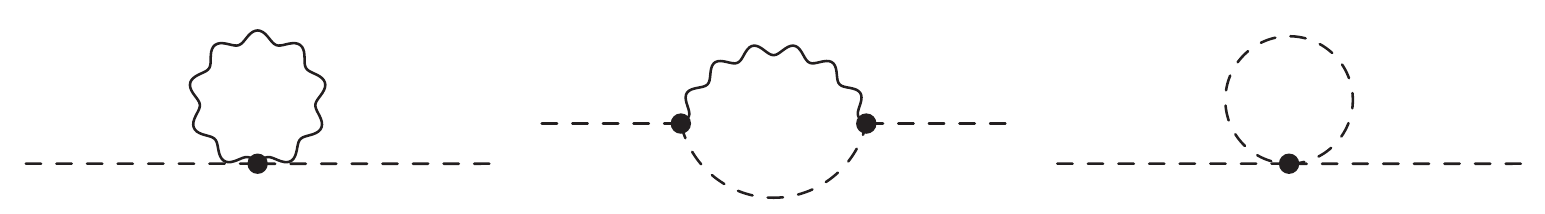}
    \caption{One-loop self-energy diagrams. Here, dashed and wavy lines represent scalar and gauge fields, respectively. \label{fig:self}}
\end{figure}

It is straightforward to compute the anomalous dimension for a scalar field using the Feynman diagram approach. This requires the calculation of the three self-energy diagrams shown in Fig.~\ref{fig:self}.

The wavefunction renormalization constant that we get after computing the loop integrals is \cite{Kang:1974yj, Andreassen:2014eha}
\be \label{eq:Ano_Dim_Diag}
    Z_{\phi} = 1 + \frac{1}{16 \pi^2} \frac{1}{\epsilon} (3 - \xi) e^2 \;,
\ee
and the field anomalous dimension is 
\be \label{eq:gamma-phi}
    \gamma_{\phi} = \frac{1}{16\pi^2} e^2 (\xi - 3) \;.
\ee
This $ \gamma_{\phi}$, along with the $\xi$ dependent effective potential computed in the functional method\footnote{It is worth mentioning that this $Z_\phi$ does not reproduce the gauge-invariant $\beta$ function if one uses the gauge-invariant effective potential calculated in \cite{Balui:2025kat}. This $\xi$-dependence is further reflected in the running of other parameters at two-loop or beyond \cite{DiLuzio:2014bua}.}~\cite{Dolan:1974gu} leads to a gauge-invariant $\beta$ function \cite{Kang:1974yj, Andreassen:2014eha}
\be \label{eq:beta_diag}
    \beta_\lambda = \frac{1}{16\pi^2} \left[{-12\lambda e^2} + \frac{10}{3} \lambda^2 + 36e^4\right] \;.
\ee
As this follows the standard approach to compute anomalous dimensions, this is the correct result for the RGE flow of the coupling $\lambda$ (see also~\cite{Kang:1974yj, Andreassen:2014eha}). 

Clearly, there is a mismatch between the mixed scalar-gauge field contribution $(\mathcal{O}(e^2\lambda))$ when comparing Eqs.~\eqref{eq:beta_diag} and ~\eqref{eq:naive}. This is also a clear example of how results from diagrammatic calculations are borrowed to obtain correct results when applying the BFM naively, as alluded to in the introduction, Sec.~\ref{sec:intro}. We will identify and rectify the reasons for this in the next section.

\subsubsection{Anomalous Dimension Contribution from Open and Closed Derivatives} \label{subsec:Ano_Dim_Open_Closed_Der}
During computation of the effective potential, we rely on the elliptic operator given in Eq.~(\ref{eq:MSQEDEO}). In the process, we focus on the terms that are quadratic in quantum fluctuations and ignore terms where derivatives act on the quantum fields, e.g., $(\hat D_\mu \eta_1) \eta^\mu \hat \phi_2$. These terms represent the mixing of the background and fluctuations associated with the scalar fields. When constructing the elliptic operator with this additional term, we encounter both open and closed derivatives once we take functional derivatives with respect to $\eta_{1, 2}$ and $\eta_{\mu, \nu}$. As our primary focus is to capture the effects of the mixing between $\hat \phi_{1,2}$ and $\eta_{1,2}$, we perform an integration by parts (IBP) as follows \footnote{This step is not mandatory. We note that if we make sure that all open derivatives emerge from terms like $(\hat{D}_\mu \eta^\mu)$, we can perform the re-summation of an infinite series of related terms, ensuring there is no loss of information relevant to the RGE computation. That is why we can perform this IBP without loss of generality.}
\be \label{eq:int-IBP}
    (\hat{D}_\mu \eta_1) \eta^\mu \hat\phi_2 \xrightarrow{\text{IBP}} - \eta_1 \hat{D}_\mu (\eta^\mu \hat\phi_2) = - \eta_1 (\hat{D}_\mu \eta^\mu) \hat\phi_2 - \eta_1 \eta^\mu (\hat{D}_\mu \hat\phi_2)\;.
\ee
Now, if we perform the functional derivative with respect to $\eta_1$ and $\eta^\mu$, we find two different operator-structures: $-(\hat D_\mu \hat \phi_2)$ and $- \hat \phi_2 \hat D_\mu$. 
The structure $(D_\mu \phi)=[D_\mu, \phi]$ is a closed derivative. Its expansion leads to terms like $D_\mu \phi,\phi D_\mu$, which are open derivatives. In the BFM, in the presence of derivative interactions, the field decompositions yield both open and closed derivatives. The closed derivatives respect the cyclic property within the trace, unlike the open derivatives. Therefore, open derivatives can lead to incomplete results. However, a careful implementation of the HK method enables us to capture the missing piece of the `naive' BFM method. In the follow-up section, we discuss how the additional contribution can be identified and how its inclusion in the anomalous dimension of the background field leads to the correct RGEs.

The elliptic operator defined earlier in Eq.~\eqref{eq:MSQEDEO} is modified through the additional term of the form Eq.~(\ref{eq:int-IBP}) present in the background field-expanded Lagrangian, see Eq.~(\ref{eq:MSQEDSL}). This term is usually ignored because it does not play any role in the effective potential computation. Yet, it turns out to be crucial for obtaining the correct anomalous dimension of the background field. The modified elliptic operator now reads
\be \label{eq:OpenDelta}
    \Delta = 
    \begin{pNiceMatrix}
        \hspace{0.3 cm} \hat{D}^2 + M^2 + \frac{\lambda}{2} \hat\phi_1^2 + \frac{\lambda}{6} \hat\phi_2^2 \hspace{0.3 cm}
        &
        \frac{\lambda}{3} \hat\phi_1 \hat\phi_2 
        &
        - 2e (\hat{D}_\nu \hat\phi_2) - e \hat\phi_2 \hat{D}_\nu
        \\[0.2cm]
        \hdottedline
        \noalign{\vskip 0.2cm}
        \hspace{0.3 cm} \frac{\lambda}{3} \hat\phi_2 \hat\phi_1
        \hspace{0.3 cm}
        &
        \hspace{0.25 cm}
        \hat{D}^2 + M^2 + \frac{\lambda}{2} \hat\phi_2^2 + \frac{\lambda}{6} \hat\phi_1^2 
        \hspace{0.25 cm}
        &
        2e (\hat D_\nu \hat \phi_1) + e \hat \phi_1 \hat D_\nu
        \\[0.2cm]
        \hdottedline
        \noalign{\vskip 0.2 cm}
        \begin{aligned}[c]
            - 2e (\hat{D}_\mu \hat\phi_2) - e \hat\phi_2 \hat{D}_\mu
        \end{aligned}
        &
        \begin{aligned}[c]
            2e(\hat{D}_\mu \hat\phi_1) + e \hat\phi_1 \hat{D}_\mu
        \end{aligned}
        &
        \begin{aligned}[c]
            & 
            \hspace{0.25 cm} - (\hat{D}_G^2 + M^2 + e^2 \hat\phi^2) g_{\mu \nu} \hspace{0.3 cm} \\
            & 
            \hspace{0.3 cm} + \left(1 - {\xi}^{-1}\right) \hat{D}^G_\mu \hat{D}^G_\nu \hspace{0.3 cm}
        \end{aligned}
        \CodeAfter
        \tikz \draw[dashed] (1-|2) -- (4-|2);
        \tikz \draw[dashed] (1-|3) -- (4-|3);        
    \end{pNiceMatrix}\;.
\ee
Here, we cannot follow the standard HK method, as $\mathcal{U}$ now contains derivatives unlike before. We define the HK in Fourier space \cite{Banerjee:2023xak} as a $t$-ordered ($\mathcal{T}$) integral 
\bea
    \tr K(t, x, x, \Delta) 
    &=& 
        \langle x | e^{- M^2 t} \; \mathcal{T} \exp \Big[-\int_0^t (D^2 + e^{M^2 t'} U e^{- M^2 t'}) \; dt'\Big] | x \rangle \\
    &=& 
        \int \frac{d^d p}{(2\pi)^d} \; \langle x |e^{- M^2 t} \mathcal{T} \exp \Big[- \int_0^t (D^2 + e^{M^2 t'} U e^{- M^2 t'}) \; dt'\Big] | p \rangle \langle p | x \rangle \n \\
    &=& 
        \int \frac{d^d p}{(2\pi)^d} \; \langle x | e^{- M^2 t} \mathcal{T} \exp \Big[- \int_0^t (D^2 + e^{M^2 t'} U e^{- M^2 t'}) \; dt'\Big] | p \rangle e^{ipx} \n \\
    &=& 
        \int \frac{d^d p}{(2\pi)^d } \; e^{- M^2 t} e^{p^2 t} \; \mathcal{T} \exp \Big[- \int_0^t (D^2 + 2i \, p \cdot D + e^{M^2 t'} U e^{- M^2 t'}) \; dt'\Big] \;. \n
\eea
Then, we define the following function 
\be
    \mathcal{F} = \mathcal{T} \, \exp \Big[- \int_0^t \mathcal{A}(s) ds \Big] = \Big[1 + \sum_{n = 1}^\infty (- 1)^n f_n (t, \mathcal{A})\Big] \;, 
\ee
and relying on that, we can express the one-loop effective Lagrangian 
\be \label{eq:Lwithf}
   \L_{\text{eff}} = c_s \tr \int_0^\infty \frac{dt}{t} \int \frac{d^d p}{(2\pi)^d t^{d/2}} e^{p^2} e^{- M^2 t} \Big[1 + \sum_{n = 1}^\infty (- 1)^n f_n (t, \mathcal{A})\Big] \;, 
\ee
with
\be
    f_n (t, \mathcal{A}) = \int_0^t ds_1 \int_0^{s_1} ds_2 \cdots \int_0^{s_{n - 1}} ds_n \, \mathcal{A} (s_1) \mathcal{A} (s_2) \cdots \mathcal{A} (s_n) \;. 
\ee
Here, $\exp\{p^2\} = \exp\{- (p_1^2 + p_2^2 + p_3^2 + p_4^2)\}$ has the usual Gaussian form.
After taking the mass matrix outside and performing the Fourier method and redefining the momentum variable, one arrives at the one-loop effective Lagrangian Eq.~(\ref{eq:Lwithf}) written as an infinite series in $f_n (t, \mathcal{A})$. Remembering that $g_{\mu \nu} = -\delta_{\mu \nu}$, the mass matrix and Gaussian variable matrix takes the form
\[\exp(-M^2 t') = 
    \begin{pmatrix}
        e^{-M^2 t'} \delta_{ab} 
        & 
        0 
        \\
        0 
        &
        e^{-M^2 t'} \delta_{\mu \nu}
    \end{pmatrix} \;, 
\quad 
\exp(p^2) = 
    \begin{pmatrix}
        e^{p^2}\delta_{ab} 
        &
        0 
        \\
        0 
        &
        e^{p^2} \delta_{\mu \nu}
    \end{pmatrix} \;. 
\]
For the case of massive scalar QED, the $\mathcal{A}$ matrix takes the following form
\be 
    \mathcal{A} =
    \begin{pNiceMatrix}
        \begin{aligned}[c]
            & 
            \hspace{0.5 cm} \hat{D}^2 + \dfrac{2ip \!\cdot\! \hat{D}}{\sqrt{t}} \\
            &
            + \dfrac{\lambda}{2} \hat\phi_1^2 + \dfrac{\lambda}{6} \hat\phi_2^2 
        \end{aligned}
        &
        {\dfrac{\lambda}{3} \hat\phi_1 \hat\phi_2}
        &
        \begin{aligned}[c]
            &
            - 2e (\hat{D}_\nu \hat\phi_2) \\
            &
            - e \hat\phi_2 \hat{D}_\nu - \dfrac{i e}{\sqrt{t}}\hat \phi_2 p_\nu
        \end{aligned}
        \\[0.8 cm] 
        \hdottedline
        \\[-0.25 cm]
        \dfrac{\lambda}{3} \hat\phi_1 \hat\phi_2
        &
        \hspace{0.25 cm}
        \begin{aligned}[c]
            & 
            \hspace{0.5 cm} \hat{D}^2 + \dfrac{2ip \!\cdot\! \hat{D}}{\sqrt{t}} \\
            & 
            + \dfrac{\lambda}{2} \hat\phi_2^2 + \dfrac{\lambda}{6} \hat\phi_1^2 
        \end{aligned}
        \hspace{0.25 cm}
        &
        \begin{aligned}[c]
            & 
            2e (\hat{D}_\nu \hat\phi_1) \\
            &
            + e \hat\phi_1 \hat{D}_\nu + \dfrac{ie}{\sqrt{t}} \hat\phi_1 p_\nu
        \end{aligned}
        \\[0.8 cm]
        \hdottedline
        \\[-0.25 cm]
        \begin{aligned}[c]
            &
            - 2e (\hat{D}_\nu \hat\phi_2) \\
            &
            - e \hat\phi_2 \hat{D}_\nu - \dfrac{i e}{\sqrt{t}}\hat \phi_2 p_\nu
        \end{aligned}
        \hspace{0.25 cm}
        &
        \hspace{0.25 cm}
        \begin{aligned}[c]
            & 
            2e (\hat{D}_\nu \hat\phi_1) \\
            &
            + e \hat\phi_1 \hat{D}_\nu + \dfrac{ie}{\sqrt{t}} \hat\phi_1 p_\nu
        \end{aligned}
        \hspace{0.25 cm}
        &
        \hspace{0.25 cm}
        \begin{aligned}[c]
            - \Big(\hat{D}_G^{2} + \dfrac{2ip \!\cdot\! \hat{D}_G}{\sqrt{t}} + e^{2} \hat\phi^{2}\Big) g_{\mu \nu} \\ 
            + \Big(1 - {\xi}^{-1}\Big) \Big(\hat{D}^G_\mu \hat{D}^G_\nu\hspace{1cm} \\
            +\dfrac{i(p_\mu \hat{D}^G_\nu +p_\nu \hat{D}^G_\mu)}{\sqrt{t}} - \dfrac{p_\mu p_\nu}{t}\Big)
        \end{aligned}
        \CodeAfter
        \tikz \draw[dashed] (1-|2) -- (6-|2);
        \tikz \draw[dashed] (1-|3) -- (6-|3);
    \end{pNiceMatrix} \;,
\ee
Open derivatives require special attention because they lead to an infinite series of relevant terms. Thus, that series needs to be resummed to capture the full contribution. 
This leads to the divergent part relevant for the anomalous dimension 
\bea
    \mathcal{S}|^{\delta Z_{\hat\phi}} 
    &=&  
        \int d^4 x \Bigg[ \Bigg. \frac{1}{64\pi^2} \frac{1}{\epsilon} \left\{ 8 - 2\left( 1 - \xi^{-1} \right) +2\left( 1 - \xi^{-1} \right)^2 - \dots\right\} e^2(\hat{D}_\mu \hat{\phi}_a)(\hat{D}_\mu \hat{\phi}_a) 
        \\ \n
    && 
        + \frac{1}{64\pi^2} \frac{1}{\epsilon} (e^2\lambda \hat{\phi}^4)\left\{ -\frac{1}{3} + \frac{1}{3} \left( 1 - \xi^{-1} \right)  - \frac{1}{3}\left(1 - \xi^{-1}\right)^2  +\dots \right\} \Bigg. \Bigg]
        \\ \n
    &=& 
        \frac{1}{64\pi^2} \frac{1}{\epsilon} \int d^4 x \Bigg[ \textcolor{red}{6 e^2(\hat{D}_\mu \hat{\phi}_a)(\hat{D}_\mu \hat{\phi}_a)} + 
        \tikz[baseline]{\node[draw, rounded corners, fill = Green!15, thick, inner sep = 3pt]
        {
        $\left(2e^2(\hat{D}_\mu\hat{\phi}_a)(\hat{D}_\mu\hat{\phi}_a) 
        - \frac{1}{3}(e^2\lambda\hat{\phi}^4)\right)$};} 
        \\ \n
    &&
        - \left(1 - \xi^{-1}\right) \tikz[baseline]{\node[draw, rounded corners, fill = cyan!15, thick, inner sep = 3 pt]
        {
        $\left(2e^2(\hat{D}_\mu\hat{\phi}_a)(\hat{D}_\mu\hat{\phi}_a) 
        - \frac{1}{3}(e^2\lambda\hat{\phi}^4)\right)$};} 
        \n \\
    &&
        + \left(1 - \xi^{-1}\right)^2 \tikz[baseline]{\node[draw, rounded corners, fill = orange!15, thick, inner sep = 3 pt]
        {
        $\left(2e^2(\hat{D}_\mu\hat{\phi}_a)(\hat{D}_\mu\hat{\phi}_a) 
        - \frac{1}{3}(e^2\lambda\hat{\phi}^4)\right)$};} - \dots\Bigg]\;.
\eea
Using the equation of motion of the background field $\mathcal{O}(\hat{\phi}^4)$
\bea \label{eq:bgeom}
     2e^2 (\hat D_\mu \hat \phi_a)(\hat D^\mu \hat \phi_a) - \frac{\lambda e^2}{3}\hat \phi^4 
     &=& 
     0 \Rightarrow  -2 \hat \phi_a \underbrace{\left( \Box + \frac{\lambda}{6}\hat \phi^2 \right)\hat \phi_a}_{=0~{\text{(EoM)}}} 
     =
     0\;,
\eea
ensures that each term in the brackets vanishes at the $\mathcal{O}(\hat{\phi}^4)$. This cancellation occurs at every order of expansion in the gauge parameter. The  wave function renormalization constant for the scalar field reads
\vspace{0.01cm}
\be
    Z_{\hat\phi} = 1 + \frac{1}{16\pi^2} \frac{3e^2}{\epsilon} \;,
\ee
which is $\xi $-independent.
After taking care of all contributions and suitably resumming them, we find the relevant finite and counterterms (in $\overline{\text{MS}}$ renormalization scheme) as follows \vspace{0.01cm}
\bea \label{eq:MSQED_eff_Act_onshell}
    \mathcal{L}^{\text{relevant}} 
    &=&
        - \frac{1}{64\pi^2} \left(6e^2 (\hat D_\mu \hat \phi_a) (\hat D_\mu \hat \phi_a) + \frac{5}{18}\lambda^2 \hat \phi^4 + 3e^4 \hat \phi^4 + \frac{e^2}{3} (\hat{F}_{\mu \nu})^2 \right) \log \left(\frac{M^2}{\mu^2}\right)
        \n \\ 
    & \quad &
       + \frac{1}{64\pi^2} \frac{1}{\epsilon} \Big(\underbrace{6\, e^2 (\hat{D}_\mu \hat\phi_a) (\hat{D}_\mu \hat \phi_a)}_{=\delta Z_{\hat\phi}} + \underbrace{\frac{5}{18}\lambda^2 \hat\phi^4 + 3e^4 \hat\phi^4}_{= \delta Z_\lambda} + \underbrace{\frac{e^2}{3} (\hat{F}_{\mu \nu})^2 }_{=\delta Z_{\hat{A}}}\Big)\;  .
\eea
\vspace{0.01cm}
Following the convention of Eqs.~\eqref{eq:ren_coef} and \eqref{eq:ren_coef_div}, we get from Eq.~\eqref{eq:MSQED_eff_Act_onshell}
    \be \label{eq: anom-dim-SQED}
    \delta Z_\lambda = \dfrac{1}{16\pi^2} \left[ \dfrac{5}{3}\lambda + 18 \dfrac{e^4}{\lambda} \right]\;, \quad \delta Z_{\hat \phi} = \dfrac{3 e^2}{16\pi^2} \;.
\ee
With Eq.~\eqref{eq:beta_form} the $\beta_\lambda$ now follows as
\vspace{0.01cm}
\bea
    \beta_\lambda 
    &=& 
        (-2) \left\{2 \frac{\d \delta Z_{\hat\phi}}{\d \lambda} \lambda + \frac{\d \delta Z_{\hat\phi}}{\d e} e \right\} \lambda + \lambda \left\{ 2 \frac{\d \delta Z_\lambda}{\d\lambda} \lambda + e \frac{\d \delta Z_\lambda}{\d e}\right\} \n\\
    &=& 
        \frac{1}{16\pi^2} \left[\frac{10}{3}\lambda^2 + 36e^4 - 12\lambda e^2\right] \;.
\eea
This is consistent with $\beta_\lambda$ computed using combined diagrammatic and functional prescriptions \cite{Kang:1974yj, Andreassen:2014eha}. The RGE for the running charge is unmodified. Using gauge-invariant $Z_{\hat \phi}$, see Eq.~\eqref{eq: anom-dim-SQED}, we compute the gauge-invariant background field anomalous dimension of $\hat \phi$ as 
\be
    \gamma_{\hat \phi} = \frac{1}{2 } \frac{\d \ln Z_{\hat\phi}}{\d \ln \mu} = - \frac{3e^2}{16 \pi^2}\;.
\ee
It is worth highlighting that, given the strategy presented here, the anomalous dimension is directly gauge parameter-independent, cf.~\cite{Kang:1974yj,Andreassen:2014eha}. Spurious gauge-parameter dependencies, see Eq.~\eqref{eq:gamma-phi}, are therefore avoided from the outset.
\subsection{More on Fermions: Yukawa Interactions}\label{sec: Yukawa}
We now turn to a toy Yukawa theory to further build support for our formalism by investigating heavy fermions in more detail. We consider a Lagrangian 
\be \label{eq:Lag_yukawa}
    \L_{(M)} = \Bar{\psi} \left( i \gamma^\mu D_\mu - M_f - y \phi \right) \psi + \frac{1}{2} (D_\mu \phi) (D^\mu \phi) - \frac{1}{2} M^2 \phi^2 - \frac{\lambda}{24} \phi^4 \;.
\ee
After performing the Wick rotation, we get the following Euclidean Lagrangian 
\be
    \L_{(E)} = - \Bar{\psi} \left( i \gamma_\mu^{(E)} D_\mu^{(E)} - M_f - y \phi \right) \psi + \frac{1}{2} \phi D^2 \phi + \frac{1}{2} M^2 \phi^2 + \frac{\lambda}{24} \phi^4 \;.
\ee
Now onward, we will suppress the index $(E)$. Following the same procedure detailed in Sec.~\ref{subsec:MSQEDQL}, we identify the elliptic operator in the basis $V^{T} = \left( \psi, \eta \right)$ as 
\be
    \Delta = 
    \begin{pmatrix}
        - i \hat {\slashed{D}} + M_f + y \hat \phi
        & 
        \quad y \hat \psi 
        \\
        - y \Bar{\hat\psi}
        &
        \quad \frac{1}{2} \left( \hat D^2 + M^2 + \frac{\lambda}{2} \hat \phi^2 \right)
    \end{pmatrix} \;.
\ee 
This can be written as 
\be
    \Delta|_{\text{diag}} = 
    \begin{pmatrix}
        - i \slashed{D} + M_f + y \hat \phi
        & \quad 0
        \\
        0 
        & \quad \frac{1}{2} \left( \hat D^2 + M^2 + \frac{\lambda}{2} \hat \phi^2 \right) - \left( -y \Bar{\hat\psi} \right) \left( - i \hat{\slashed{D}} + M_f + y \hat \phi \right)^{- 1} \left( y \hat \psi \right) 
    \end{pmatrix} \;,
\ee
after diagonlization and, we can simplify its scalar part
\bea
    \Delta|_{\text{diag}}^{(S)} = 
    && 
        \frac{1}{2} \left( \hat D^2 + M^2 + \frac{\lambda}{2} \hat \phi^2 \right) - \left( -y \Bar{\hat\psi} \right) \left( - i \hat{\slashed{D}} + M_f + y \hat \phi \right)^{- 1} \left( y \hat \psi \right) \n \\
    \xrightarrow{M^2_f\gg \hat{D}^2 
    } &&
        \frac{1}{2} \left( \hat D^2 + M^2 \right) + \frac{1}{2} \left( \frac{\lambda}{2} \hat \phi^2 + \dfrac{2 \, y^2}{M_f} \Bar{\hat\psi}  \hat \psi - \dfrac{2 \, y^3}{M_f^2} \hat \phi \Bar{\hat\psi}  \hat \psi \; + \dfrac{2 \, y^2}{M_f^2} \left( \Bar{\hat\psi} \; i \hat{\slashed{D}} \hat \psi \right) \right) \;. \n 
\eea
This leads to the the final expression (cf. Eq.~\eqref{eq:ell}), identifying the scalar contribution to $U$
\be
    U_S = \frac{\lambda}{2} \hat \phi^2 + \dfrac{2 \, y^2}{M_f} \Bar{\hat\psi}  \hat \psi - \dfrac{2 \, y^3}{M_f^2} \hat \phi \Bar{\hat\psi}  \hat \psi \; + \dfrac{2 \, y^2}{M_f^2} \left( \Bar{\hat\psi} \; i \hat{\slashed{D}} \hat \psi \right) \;.
\ee
Following the HK treatment for fermions detailed in \cite{Chakrabortty:2023yke}, we obtain the fermionic contribution to the $U$ operator
\be
    U_f = y^2 \hat \phi^2 - y \; i \hat{\slashed{D}} \hat \phi  + 2 M_f  y  \hat \phi \;.
\ee
Now the interaction of this theory is captured in $\mathcal{U}$, which can be written as 
\be
    \mathcal{U} = 
    \begin{pmatrix}
        U_f
        & \quad
        0
        \\
        \
        0 
        & \quad 
        U_S
    \end{pmatrix}\;.
\ee
Following the expression of the effective Lagrangian in Eq.~\eqref{eq:Lwithf}, we get the following relevant contributions to the counterterms as
\bea
 \L_{\text{CT}}|_Y &=& \Big\{ \frac{1}{32 \pi^2} \frac{1}{\epsilon} \frac{M^2}{M_f^2}   \Big[ \underbrace{-2 \, y^3 \hat \phi \Bar{\hat\psi} \hat \psi}_{=Z_y} + \underbrace{2 \, y^2 \Bar{\hat\psi} \; i \hat{\slashed{D}} \hat \psi}_{=Z_{\hat \psi}} \Big]
 +  \frac{1}{64 \pi^2}  \frac{1}{\epsilon} \underbrace{\frac{\lambda^2 \hat \phi^4}{4}}_{=Z_\lambda} \Big\}_{\text{scalar}} \n \\
 &+ & \Big\{- \frac{1}{64 \pi^2}  \frac{1}{\epsilon} \Big[ \underbrace{4 \, y^4 \hat \phi^4}_{=Z_\lambda} - \underbrace{4 \, y^2 (\hat D_\mu \hat \phi) (\hat D_\mu \hat \phi)}_{=Z_{\hat \phi}} \Big] \Big\}_{\text{fermion}}\;,
\eea
which leads to the renormalization constants in the heavy fermion mass limit
\be
\delta Z_{\hat \phi} = \frac{2y^2}{16\pi^2}\;,
\quad \delta Z_\lambda = \frac{1}{16\pi^2} \left[ -\frac{3\lambda}{2} + \frac{24y^4}{\lambda} \right]\;, 
\quad \delta Z_{\hat \psi} = \frac{2y^2}{32\pi^2} \frac{M^2}{M_f^2}\;, 
\quad \delta Z_y = \frac{2y^2}{32\pi^2} \frac{M^2}{M_f^2}\;.
\ee
These results agree with the diagrammatic approach, which is relegated to Appendix~\ref{app:fermion-diag}.

\section{Conclusions}
Calculations in the realm of effective field theory and their perturbative improvements have led to considerable progress in formulating new approaches to computing phenomenologically relevant quantities through functional rather than diagrammatic techniques~\cite{Henning:2014wua,Drozd:2015rsp,Henning:2016lyp,Dittmaier:2021fls}. In this context, the Background Field Method, along with Background Gauge Fixing, adds many desirable theoretical properties, such as manifest gauge-invariance expressed in its naive, classical meaning. When capturing effects beyond the semi-classical level, algebraic techniques can fail to yield correct results if limited to the background-field dynamics. This is not a surprise, but merely an expression of the re-organization of the BFM when compared to the standard diagrammatic approach. Concretely, a computation of the RGE evolution of Lagrangian parameters at one-loop order requires the inclusion of BFM contributions beyond the classical background sources; only once these are provided by `subsidiary' calculations, the correct quantum dynamics can be recovered (see~\cite{Denner:1994xt} for BFM applications in the SM). This creates a technical bottleneck, especially when algebraic techniques are employed to construct effective theories. 

In this work, for the first time, we propose an improved BFM blended with the Heat Kernel method to capture the mixing of background and quantum fluctuation dynamics. This enables us to compute gauge-invariant and gauge parameter-independent (in case of gauge theory) anomalous dimensions of fields, and beta functions of the Lagrangian parameters without borrowing any result from diagrammatic or other methods, when background fields are on-shell. We highlight the importance of open derivatives (ignored earlier) through integration-by-parts and resummation to achieve our goal. We validate our method for the Yukawa interaction and the electroweak sector of the SM, by computing the gauge-invariant $\beta$ functions, in agreement with the diagrammatic methods. Gauge-independence of all quantities at all steps provides a promising and economic avenue for future applications.

\section*{Acknowledgments}
J.C. acknowledges the hospitality of HRI, Allahabad, India, during a long-term visit and the SANGAM-2026 conference, where part of the research was done. We acknowledge the engaging and insightful discussions with Sabyasachi Chakraborty, Goutam Das, Debmalya Dey, and Apratim Kaviraj. This work is supported by the Core Research Grant (CRG/2023/003200), SERB, India.

\appendix
\section{Bosonic RGEs of the SM}
Extending the Abelian Higgs model to the more realistic case of the Standard Model, we turn to the SM Lagrangian 
\begin{equation} \label{eq:SMLag1}
\mathcal{L}_{\rm{SM}} = \mathcal{L}_{\rm{YM}} + \mathcal{L}_{\rm{H}} + \mathcal{L}^{\rm{Fermi}}_{\rm{g.f.}} \;  .
\end{equation} 
Here, we focus on the electroweak sector $(SU(2)_L \times U(1)_Y)$, and the relevant part of the Lagrangian is depicted as in \cite{Balui:2025yvd}:
\begin{eqnarray}\label{eq:SMLag2}
\mathcal{L}_{\rm{YM}} &= & 
-\frac{1}{4} \left(\partial_\mu W^a_\nu - \partial_\nu W^a_\mu + g \epsilon^{abc} W^b_\mu W^c_\nu \right)^2
-\frac{1}{4} \left(\partial_\mu B_\nu - \partial_\nu B_\mu \right)^2  , \n \\
\mathcal{L}_{\rm{H}} &=& \left( D_\mu H \right)^\dagger \left( D^\mu H \right) - V(H)  \;, 
\end{eqnarray}
where the covariant derivative is $D_\mu \equiv \partial_\mu - i g \frac{\sigma^a}{2} W^a_\mu + i g' \frac{Y}{2} B_\mu$, and the scalar potential is $V (H) = m^2 H^\dag H + \lambda (H^\dag H)^2$. The SM gauge fields are expressed in terms of background fields and quantum fluctuations as follows:  $W_\mu = \hat W^\mu + \eta_{W}^{\mu}$ and $B_\mu = \hat B_\mu + \eta_{B}^\mu$, where the Higgs doublet is given as 
\be
    H(x) \equiv
    \frac{1}{\sqrt{2}} 
    \begin{pmatrix}
        \eta_1(x) + i\eta_2(x)
        \\
        \hat \phi(x) + h(x) + i\eta_3(x)
    \end{pmatrix}\;.
\ee
Here, the hat-fields are backgrounds and $\eta$'s are the associated quantum fluctuations. We assume the non-zero background is associated only with one of the real scalar components\footnote{The conclusion remains unaltered even if all the real scalar components possess non-zero background fields due to an $SO(4)$ symmetry of the unbroken SM scalar sector.} The gauge fixing term in the BGF method is written as in \cite{Schwartz:2014sze,Sarkar:1974db,Sarkar:1974ni} 
\be
\mathcal{L}^{\rm{Fermi}}_{\rm{g.f.}}  =  -\frac{1}{2 \xi_W} \left( \hat D^{G1}_\mu \eta_W^\mu \right)^2 
-\frac{1}{2 \xi_B} \left( \hat D^{G2}_\mu \eta_B^\mu \right)^2\;,
\ee
where the covariant derivative acting on the gauge field is $\hat D^{G1}_\mu \eta^{\mu,a}_W = \partial_\mu \eta^{\mu,a}_W + g \epsilon^{abc} \hat W_\mu^b \eta^{\mu,c},\, \hat D^{G2}_\mu \eta^{\mu}_B = \partial_\mu \eta^{\mu}_B$. It is worth mentioning that for this choice of background Fermi gauge which is independent of $\hat{\phi}$,  the ghost fields are massless. Thus, the effective potential and the RGEs that depend on the ghost mass, do not receive any contribution from the ghost sector \cite{DiLuzio:2014bua}. Here, notably the gauge fields, which have field dependent masses, have three polarisation in this gauge. 
Following the same procedure discussed in Sec.~\ref{subsec:Ano_Dim_Open_Closed_Der} we get the following form for $\mathcal{A}_{\text{SM}}$
\be
    \mathcal{A}_{\text{SM}} = 
    \begin{pmatrix}
        \mathcal{A}_{\text{SM}}|_{\text{scalar}} & \qquad \mathcal{A}_{\text{SM}}|_{\text{int}}
        \\
        (\mathcal{A}_{\text{SM}}|_{\text{int}})^{T} & \qquad (\mathcal{A}_{\text{SM}}|_{\text{gauge}})
    \end{pmatrix}\;,
\ee
where 
\be
    \mathcal{A}_{\text{SM}}|_{\text{scalar}} = 
    \begin{pNiceMatrix}
        \begin{aligned}[c]
        \hat D^2 + \dfrac{2 i p \cdot \hat D}{\sqrt{t}} \hspace{0.2cm} \\+ \lambda \hat \phi^2 \hspace{0.2cm}\\[0.2cm]
        \end{aligned}
        & 0 & 0 & 0 
        \\[-0.3cm]
        \hdottedline
        \\
        0 & 
        \begin{aligned}[c]        
        \hspace{0.2cm}\hat D^2 + \dfrac{2 i p \cdot \hat D}{\sqrt{t}} \hspace{0.2cm}\\+ \lambda \hat \phi^2 \hspace{0.2cm}
        \\[0.2cm]
        \end{aligned}
        & 0 & 0
        \\[-0.3cm]
        \hdottedline
        \\
        0 & 0 & 
        \begin{aligned}[c]
            \hspace{0.2 cm} \hat{D}^2 + \dfrac{2ip \!\cdot\! \hat{D}}{\sqrt{t}} \\
            + 3\lambda \hat\phi^2 \hspace{0.2cm} \\[0.2cm]
        \end{aligned}
        & 0
        \\[-0.3cm]
        \hdottedline
        \\
        0 & 0 & 0 & 
        \begin{aligned}[c]
            \hspace{0.2cm}\hat D^2 + \dfrac{2 i p \cdot \hat D}{\sqrt{t}} \\
            + \lambda \hat \phi^2
        \end{aligned}
        \CodeAfter
        \tikz \draw[dashed](1-|2) -- (8-|2);
        \tikz \draw[dashed](1-|3) -- (8-|3);        
        \tikz \draw[dashed](1-|4) -- (8-|4);        
    \end{pNiceMatrix}\;,
\ee

We have also shown explicitly for the SM, similar to the SQED case, that with arbitrary $\xi$, we can resum the infinite series to ensure the absence of the gauge parameter in the effective Lagrangian once we employ the background field EOM $( \hat{\phi}\Box \hat{\phi}+\lambda \hat{\phi}^4)=0$ at the $\mathcal{O}(\hat \phi^4)$. This, suggests that for complicated gauge theories it is also safe to choose the gauge parameters $\xi=1$. This choice reduces the elliptic operator to a minimal one, and the computation of an effective Lagrangian is more straightforward. 
\be
\mathcal{A}_{\text{SM}}|_{\text{gauge}} =
- 
\begin{pNiceMatrix}
	\begin{aligned}[c]
		& \hspace{0.1cm} \bigg[\hat D_{G1}^2 
		+ \frac{2 i p \cdot \hat D_{G1}}{\sqrt{t}}\hspace{0.1cm}  \\
		&\quad + \frac{1}{4} g^2 \hat \phi^2\bigg] g_{\mu\nu} \\
		&-(1-\frac{1}{\xi_W}) D_\mu^{G1} D_\nu^{G1} \\[0.2cm]
	\end{aligned}
	&
	-2g \hat{F}_{\mu \nu;3}
	&
	2g \hat{F}_{\mu \nu; 2}
	&
	0
	\\[-0.3cm]
	\hdottedline
	\\
	2g \hat{F}_{\mu \nu;3}
	&
	\begin{aligned}[c]
		& \hspace{0.1cm} \bigg[\hat D_{G1}^2 
		+ \frac{2 i p \cdot \hat D_{G1}}{\sqrt{t}}\hspace{0.1cm}  \\
		&\quad + \frac{1}{4} g^2 \hat \phi^2\bigg] g_{\mu\nu} \\
		&-(1-\frac{1}{\xi_W}) D_\mu^{G1} D_\nu^{G1} \\[0.2cm]
	\end{aligned}
	&
	-2g \hat{F}_{\mu \nu; 1}
	&
	0
	\\[-0.3cm]
	\hdottedline
	\\
	-2g \hat{F}_{\mu \nu; 2}
	&
	2g \hat{F}_{\mu \nu; 1}
	&
	\begin{aligned}[c]
		& \hspace{0.1cm} \bigg[\hat D_{G1}^2
		+ \frac{2 i p \cdot \hat D_{G1}}{\sqrt{t}} \hspace{0.1cm} \\
		&\quad + \frac{1}{4} g^2 \hat \phi^2\bigg] g_{\mu\nu} \\
		&-(1-\frac{1}{\xi_W}) D_\mu^{G1} D_\nu^{G1} \\[0.2cm]
	\end{aligned}
	&
	[\frac{1}{4} g g' \hat \phi^2]g_{\mu\nu}
	\\[-0.3cm]
	\hdottedline
	\\
	0
	&
	0
	&
	[ \frac{1}{4} g g' \hat \phi^2]g_{\mu\nu}
	&
	\begin{aligned}[c]
		& \hspace{0.1cm} \bigg[ \hat D_{G2}^2 
		+ \frac{2 i p \cdot \hat D_{G2}}{\sqrt{t}} \\
		&\quad + \frac{1}{4} g'^2 \hat \phi^2\bigg] g_{\mu\nu} \\
		&-(1-\frac{1}{\xi_B}) D_\mu^{G2} D_\nu^{G2} \\[0.2cm]
	\end{aligned}
	\CodeAfter
	\tikz \draw[dashed] (1-|2) -- (8-|2);
	\tikz \draw[dashed] (1-|3) -- (8-|3);
	\tikz \draw[dashed] (1-|4) -- (8-|4);
\end{pNiceMatrix}\;,
\ee
\be
    \mathcal{A}_{\text{SM}}|_{\text{int}} =
    \begin{pNiceMatrix}
        0
        &
        \begin{aligned}[c]
        &- (g \hat D_\nu \,\hat \phi)
        - \frac{g}{2} \hat \phi \hat D_\nu \hspace{0.1cm} \\
        & \quad - \frac{ig p_\nu \hat\phi}{2\sqrt{t}}\\[0.2cm]
        \end{aligned}
        &
        0
        &
        0
        \\[-0.3cm]
        \hdottedline
        \\
        \begin{aligned}[c]
        &- (g \hat D_\nu \,\hat \phi)
        - \frac{g}{2}\hat\phi \hat D_\nu \hspace{0.1cm} \\
        &\quad - \frac{i g p_\nu \hat\phi}{2\sqrt{t}}\\[0.2cm]
        \end{aligned}
        &
        0
        &
        0
        &
        0
        \\[-0.4cm]
        \hdottedline
        \\
        0
        &
        0
        &
        0
        &
        0
        \\[-0.3cm]
        \hdottedline
        \\
        0
        &
        0
        &
        \begin{aligned}[c]
        &\hspace{0.1cm} (g\hat D_\nu \,\hat \phi) 
        + \frac{g}{2}\hat\phi \hat D_\nu \hspace{0.1cm} \\
        &\quad + \frac{i g p_\nu \hat\phi}{2\sqrt{t}}
        \end{aligned}
        &
        \begin{aligned}[c]
        &\hspace{0.1cm} (g' \hat D_\nu \,\hat \phi)
        + \frac{g'}{2}\hat\phi \hat D_\nu \\
        &\quad + \frac{i g' p_\nu \hat\phi}{2\sqrt{t}}\\[0.2cm]
        \end{aligned}
        \CodeAfter
        \tikz \draw[dashed] (1-|2) -- (8-|2);
        \tikz \draw[dashed] (1-|3) -- (8-|3);
        \tikz \draw[dashed] (1-|4) -- (8-|4);
    \end{pNiceMatrix}\;.
\ee
Following the previously discussed HK method, the one-loop effective Lagrangian can be given as 
\bea \label{eq:EffLb2divSMHK}
  \L_{\text{SM,eff}}^{\text{fin}} + \L_{\text{SM,eff}}^{\text{div}} && 
  \\
  &\hspace{-4 cm}= & \hspace{-2 cm}
    - \frac{1}{64\pi^2} \Bigg[\frac{3}{2} (g'^2 + 3g^2) (\hat D_\mu \hat \phi) (\hat D_\mu \hat \phi)
    + \left(12\lambda^2 + \frac{3}{16}g'^2 + \frac{3}{8}{g^2}{g'^2} + \frac{9}{16}g^4\right) {\hat \phi}^4\Bigg] \log \left(\frac{M^2}{\mu^2}\right)
    \n \\
 &\hspace{-4 cm} & \hspace{-2 cm}
    + \frac{1}{64\pi^2} \frac{1}{\epsilon} \Bigg[  
     \left(12\lambda^2 + \frac{3}{16}g'^2 + \frac{3}{8}{g^2}{g'^2} + \frac{9}{16}g^4\right) {\hat \phi}^4 \; 
 +    \Bigg\{
    \textcolor{blue}{\Big[}
   \textcolor{red} {\frac{3}{2}g'^2 (\hat D_\mu \hat \phi) (\hat D_\mu \hat \phi)} + \frac{1}{2}g'^2 \;
    \tikz[baseline]{\node[draw, rounded corners, fill = green!15, thick, inner sep = 3 pt]
    	{
    		$\biggl((\hat D_\mu \hat \phi) (\hat D_\mu \hat \phi)
    		-\lambda \hat\phi^4\biggr)$};} \n \\
 &\hspace{-4 cm} & \hspace{-2 cm}
  -\frac{1}{2}\Bigl(1-\frac{1}{\xi_B}\Bigr)g'^2 \;
    \tikz[baseline]{\node[draw, rounded corners, fill = green!15, thick, inner sep = 3 pt]
    	{
    		$\biggl((\hat D_\mu \hat \phi) (\hat D_\mu \hat \phi)
    		-\lambda \hat\phi^4\biggr)$};}
    +\frac{1}{2}\Bigl(1-\frac{1}{\xi_B}\Bigr)^2 g'^2 \;
    \tikz[baseline]{\node[draw, rounded corners, fill = green!15, thick, inner sep = 3 pt]
    	{
    		$\biggl((\hat D_\mu \hat \phi) (\hat D_\mu \hat \phi)
    		-\lambda \hat\phi^4\biggr)$};}
    +\cdots
    \textcolor{blue}{\Big]}
    \n \\[0.35 cm]
 &\hspace{-4 cm} & \hspace{-2 cm}
    +\textcolor{blue}{\Big[}
   \textcolor{red} {\frac{9}{2}g^2 (\hat D_\mu \hat \phi) (\hat D_\mu \hat \phi)}
    + \frac{3}{2}g^2 \;
    \tikz[baseline]{\node[draw, rounded corners, fill = cyan!15, thick, inner sep = 3 pt]
    	{
    		$\biggl((\hat D_\mu \hat \phi) (\hat D_\mu \hat \phi)
    		-\lambda \hat\phi^4\biggr)$};}
-\frac{3}{2}\Bigl(1-\frac{1}{\xi_W}\Bigr)g^2 \;
    \tikz[baseline]{\node[draw, rounded corners, fill = cyan!15, thick, inner sep = 3 pt]
    	{
    		$\biggl((\hat D_\mu \hat \phi) (\hat D_\mu \hat \phi)
    		-\lambda \hat\phi^4\biggr)$};}
    \n \\
    &&
    \hspace{3 cm}+\frac{3}{2}\Bigl(1-\frac{1}{\xi_W}\Bigr)^2 g^2 \;
    \tikz[baseline]{\node[draw, rounded corners, fill = cyan!15, thick, inner sep = 3 pt]
    	{
    		$\biggl((\hat D_\mu \hat \phi) (\hat D_\mu \hat \phi)
    		-\lambda \hat\phi^4\biggr)$};} +\cdots
    \textcolor{blue}{\Big]}
    \Bigg\} \Bigg]
    \;.
\eea
Here, we note that 
\bea \label{eq:SM-CTs}
    \delta Z_\lambda 
    &=& 
        \frac{1}{16\pi^2} \cdot \frac{1}{\lambda} \left[12\lambda^2 
        + \frac{3}{16}(g'^4 + 2{g^2}{g'^2} + 3{g^4})\right], 
    \\
    \delta Z_{\hat \phi} 
    &=& 
        \frac{1}{64\pi^2} \cdot 3(g'^2 + 3{g^2}) \;.
\eea
The $\beta$ function then follows as
\bea \label{eq:SM-beta-function}
    \beta_\lambda 
    &=& 
        \lambda \left\{2\frac{\d \delta Z_\lambda}{\d \lambda} \lambda + \frac{\d \delta Z_\lambda}{\d g'} g' + \frac{\d \delta Z_\lambda}{\d g} g\right\} 
        + -2\left\{2\frac{\d \delta Z_{\hat \phi}}{\d \lambda} \lambda + \frac{\d \delta Z_{\hat \phi}}{\d g'} g' + \frac{\d \delta Z_{\hat \phi}}{\d g} g\right\} \lambda 
        \n \\
     &=& 
        \frac{1}{16\pi^2} \Bigg[24\lambda^2 + \frac{3}{8}g'^4
        + \frac{3}{4}{g^2}{g'^2} + \frac{9}{8}g^4 
        - 3\lambda{g'^2} - 9\lambda{g^2}\Bigg]\;.
\eea
The background field anomalous dimension of the SM Higgs field $\hat \phi$ is computed using Eq.~\eqref{eq:SM-CTs} as
\be
    \gamma_{\hat\phi}|_{\text{SM}} = -\frac{1}{64 \pi^2}  3(g'^2 + 3{g^2})\;,
\ee
which is independent of the gauge parameter. The gauge parameter-dependent field anomalous dimension 
\be
    \gamma_{\phi}|_{\text{SM}} = \frac{1}{64 \pi^2} \Big(g'^2 (\xi_B - 3) + 3g^2 (\xi_W - 3)\Big)\;,
\ee
has been computed in Ref.~\cite{DiLuzio:2014bua}.
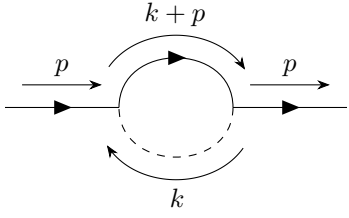
\begin{figure}[!t]
\centering
    \begin{tikzpicture}
        \begin{feynman}
        \vertex (a);
        \vertex [right = 1.5cm of a] (b) ;
        \vertex [right = 1.5cm of b] (c) ;
        \vertex [right = 1.5cm of c] (d);
        \diagram*{
            (a) -- [fermion, momentum = \(p\)] (b) -- [fermion, half left, momentum = \(k + p\)] (c) --[scalar, half left, momentum = \(k\)] (b);
            (c) -- [fermion, momentum = \(p\)] (d);
        };
        \end{feynman}           
    \end{tikzpicture}
    \caption{One-loop Self-Energy: dashed line and solid line represent scalar and fermion fields,  respectively, with Feynman Rules derived from Eq.~\eqref{eq:Lag_yukawa}. \label{fig:2pt}}
\end{figure}
\section{Diagrammatic Counterterms in the Yukawa Theory}\label{app:fermion-diag}
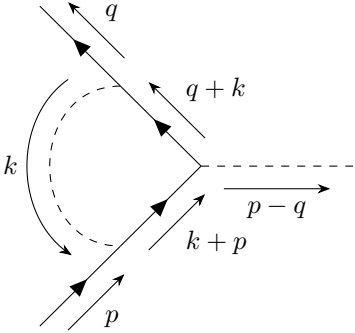
\begin{figure}[htb]
\centering
    \begin{tikzpicture}
        \begin{feynman}
        \vertex (a);
        \vertex [above right = of a] (b) ;
        \vertex [above right = of b] (c) ;
        \vertex [above left = of c] (d);
        \vertex [above left = of d] (e);
        \vertex [right = 2 cm of c] (f);
        \diagram*{
            (a) -- [fermion, momentum' = \(p\)] (b) -- [fermion, momentum' = \(k + p\)] (c) --[fermion, momentum' = \(q + k\)] (d) --[fermion, momentum' = \(q\)] (e);
            (c) -- [scalar, momentum' = \(p - q\)] (f);
            (d) --[scalar, half right, momentum' = \(k\)] (b);
        };
        \end{feynman}            
    \end{tikzpicture}
    \caption{One-loop Vertex Correction: dashed line and solid line represent scalar and fermion fields,  respectively. \label{fig:3pt}}
\end{figure}

In this section, we present the results for computing the divergent parts of the relevant two and three-point functions (Figs.~\ref{fig:2pt} and \ref{fig:3pt}, respectively) required for the renormalization of the Yukawa model using the diagrammatic approach. These follow standard techniques (see, e.g.~\cite{Denner:1991kt}) and are quoted here for completeness.
The divergent contributions in dimensional regularization of the two-point function can be extracted from translating Fig.~\ref{fig:2pt} into an algebraic expression:
\bea
    i \, \mathcal{M}_1
    = && 
        (- i y)^2 \int \frac{d^d k}{(2 \pi)^d} \, \frac{i}{k^2 - M^2 + i\varepsilon}  \frac{i}{(\slashed{k} + \slashed{p}) - M_f + i\varepsilon} \n \\
    \simeq &&
        y^2 \int \frac{d^d k}{(2 \pi)^d} \, \frac{1}{k^2 - M^2 + i\varepsilon}  \left(- \frac{1}{M_f} \right) \left(1 - \frac{\slashed{k} + \slashed{p}}{M_f} \right)^{-1} \n \\
    \supset &&
              - \frac{y^2}{M_f}\, \, \frac{- i}{(4 \pi)^2} \, M^2
              \left( - \frac{1}{\epsilon} \right) \left(1 + \frac{\slashed{p}}{M_f} \right) 
    \supset 
              -\frac{i}{16 \pi^2}\, y^2\, \slashed{p}\, \frac{M^2}{M_f^2} \frac{1}{\epsilon}\; .
\eea
Similarly, the three-point function of Fig.~\ref{fig:3pt} yields
%
\bea
    i \, \mathcal{M}_2
    = && 
        (- i y)^3 \int \frac{d^d k}{(2 \pi)^d} \, \frac{i}{k^2 - M^2 + i\varepsilon} \frac{i}{(\slashed{q} + \slashed{k}) - M_f + i\varepsilon}  \frac{i}{(\slashed{p} + \slashed{k}) - M_f + i\varepsilon} \n \\
    \simeq &&
        y^3 \int \frac{d^d k}{(2 \pi)^d} \, \frac{1}{k^2 - M^2 + i\varepsilon}  \left(- \frac{1}{M_f} \right)^{\!2} \left(1 + \frac{\slashed{q} + \slashed{k}}{M_f} \right) \left(1 + \frac{\slashed{p} + \slashed{k}}{M_f} \right) \n \\
    \supset &&
        \frac{y^3}{M_f^2} \int \frac{d^d k}{(2 \pi)^d} \, \frac{1}{k^2 - M^2 + i\varepsilon}  \left(1 + \frac{\slashed{k}\slashed{k}}{M_f^2} \right) 
    \supset 
        \frac{i}{16 \pi^2}\, y^3\, \frac{1}{\epsilon}\, \frac{M^2}{M_f^2} \; .
\eea
We can extract the following mass-dependent counter terms $\mathcal{O}(\frac{M^2}{M_f^2})$ from the above amplitudes, calculated in the limit $M_f \gg M$ as
\be
\delta Z_\psi = \frac{1}{16\pi^2}\, y^2\, \frac{M^2}{M_f^2}\;, \quad 
\delta Z_y = \frac{1}{16\pi^2}\, y^2\, \frac{M^2}{M_f^2}\;.
\ee
that match the ones computed employing the background method as demonstrated above.

\section*{Data Availability Statement}
No Data associated with the manuscript.

\bibliographystyle{jhep}
\bibliography{references}

\end{document}